\documentclass[10pt]{article} 
\usepackage{times}

\usepackage{fullpage}
\usepackage[american]{babel}

\usepackage{amsmath}  
\usepackage{amssymb}  
\usepackage{amsthm}
\usepackage{epsfig} 

\newcommand{\drop}[1]{}

\newcommand{\equals}{\stackrel{\mathrm{def}}{=}}

\newtheorem{lemma}{Lemma} 
 
\newtheorem{counterexample}{Counterexample} 
\newtheorem{definition}{\bf Definition} 
\newtheorem{validity test}{Validity Test}

\newcommand{\EDF}{\operatorname{EDF}}
\newcommand{\LLF}{\operatorname{LLF}}
\newcommand{\gEDF}{\operatorname{global-EDF}}

\newcommand{\lcm}{\operatorname{lcm}}
\newcommand{\FTP}{\operatorname{FTP}}
\newcommand{\FJP}{\operatorname{FJP}}
\newcommand{\WCET}{\operatorname{WCET}}
\newcommand{\CPU}{\operatorname{CPU}}
\begin{document}

\title{Exact Schedulability Test for global-{EDF} Scheduling of Periodic Hard Real-Time Tasks on Identical Multiprocessors}


\author{Jo\"{e}l Goossens\thanks{Brussels University, U.L.B., Brussels, Belgium.} \and Patrick Meumeu Yomsi\thanks{Postdoctoral researcher of the F.N.R.S, Belgium.}}
\date{}

\maketitle

                                                                                                         

\drop{
\begin{abstract}
In this paper we consider the scheduling problem of hard real-time systems composed of periodic constrained-deadline tasks upon {\em identical} multiprocessor platforms. We assume that tasks are scheduled by using the $\gEDF$ scheduler. We establish an {\em exact} schedulability test for this scheduler by exploiting on the one hand its predictability property and by providing on the other hand a feasibility interval so that if it is possible to find a valid schedule for all the jobs contained in this interval, then the whole system will be stamped feasible. In addition, we show by means of a  counterexample that the feasibility interval, and thus the schedulability test, proposed by Leung~\cite{Leung89} is incorrect and we show which arguments are actually incorrect. 
\end{abstract}
}


\section{Introduction}\label{Introduction}

Over the years, the preemptive periodic constrained-deadline task model~\cite{GFS03} has proven remarkably useful for the modeling of recurring processes that occur in hard real-time computer application systems, where the failure to satisfy any constraint may have disastrous consequences. The problem of scheduling such tasks upon a single processor ($\CPU$) so that all the deadlines are met has been widely studied in the literature and is now well understood. The most important point in this direction being that an {\em optimal online scheduler}, commonly known as {\em Earliest Deadline First} ($\EDF$), has been derived. $\EDF$ is a priority-based scheduler which assigns priorities to jobs so that the shorter the absolute deadline of a job the higher its priority. This scheduler is optimal with the interpretation that if a periodic constrained-deadline task system can be successfully scheduled with another scheduler upon a single $\CPU$, then it can also be successfully scheduled using $\EDF$. However, a very large number of applications nowadays turns out to be executed upon more than one $\CPU$ for practical and economic reasons due to the advent of multicore technologies. For such applications, even though $\EDF$ is no longer optimal~\cite{dhall78}, much recent work gave rise to multiple investigations and thus many alternative algorithms based on this scheduling policy have been developed due to its optimality upon uniprocessor platforms~\cite{Liu73}. Most results have been derived under either {\em global} or {\em partitioned} scheduling techniques. In {\em global} scheduling~\cite{Bertogna09}, all the tasks are stored in a single priority-ordered queue and the global scheduler selects for execution the highest priority tasks from this queue. In this framework, tasks are allowed to migrate at runtime from one $\CPU$ to another in order to complete their executions~\cite{CG06, MBTHESIS}. In {\em partitioned} scheduling~\cite{SanFis06}, all the tasks are first assigned statically to the $\CPU$s, then each $\CPU$ uses independently its local scheduler at runtime. Despite these two scheduling techniques are {\em incomparable}~\cite{Baruah2007Techniques-for-} in the sense that there are systems which are schedulable with partitioning and not by global and conversely, and despite the high number of interesting results that have already been derived up to now, many open questions still remain to be answered, especially when {\em global schedulers} are considered. Regarding this kind of schedulers, an important issue consists in deriving an {\em exact} schedulability test by exploiting on the one hand the predictability property of the scheduler and by providing on the other hand a feasibility interval so that if it is possible to find a valid schedule for all jobs contained in this interval, then the whole system will be stamped feasible. 

\paragraph{Related work.} In recent years, as most global schedulers are predictible, extensive efforts have been performed towards addressing the problem of determining a feasibility interval for the global scheduling of periodic constrained-deadline tasks upon multiprocessor platforms. That is, to derive an interval of time so that if it is possible to find a valid schedule for all jobs contained in this interval, then the whole system is feasible. Up to now, sound results have been obtained only in the particular case where tasks are scheduled by using an Fixed-Task-Priority ($\FTP$) scheduler~\cite{CG06, CG07}. Being an $\FTP$ scheduler one where all the jobs belonging to a task are assigned the same priority as the priority assigned to the task beforehand (i.e., at design time). We are not currently aware of any existing result concerning the feasibility interval for Fixed-Job-Priority ($\FJP$) schedulers in the literature, except the one proposed by Leung in~\cite{Leung89}. However, we show that this result is actually wrong. An $\FJP$ scheduler is one where two jobs belonging to the same task may be assigned different priorities.  

\paragraph{This research.} In this paper, we derive a feasibility interval for an $\FJP$ scheduler, namely $\gEDF$. To the best of our knowledge, this will be the \emph{first} valuable feasibility interval for $\FJP$ schedulers since the one proposed by Leung in~\cite{Leung89} is flawed. Based on this feasibility interval and considering the predictability property of this scheduler, our main contribution is therefore an {\em exact} schedulability test for the $\gEDF$ scheduling of periodic hard real-time tasks upon identical multiprocessor platforms.

\paragraph{Paper organization.} The remainder of this paper is structured as follows. Section~\ref{Model} presents the system model and the scheduler that are used throughout the paper. Section~\ref{Definitions and properties} provides the reader with some useful definitions and properties. Section~\ref{The proposed approach} presents our main contribution. Finally, Section~\ref{Conclusion} concludes the paper.

\section{System model}\label{Model}

Throughout this paper, all timing characteristics in our model are assumed to be non-negative integers, i.e., they
are multiples of some elementary time interval (for example the $\CPU$ tick, the smallest indivisible $\CPU$ time unit).

\subsection{Task specifications}

We consider the preemptive scheduling of a hard real-time system $\tau \equals \{\tau_1, \tau_2, \ldots, \tau_{n}\}$ composed of $n$ tasks upon $m$ {\em identical} $\CPU$s according to the following interpretations.

\begin{itemize}
\renewcommand{\labelitemi}{$\triangleright$}
\item {\em Preemptive scheduling}: an executing task may be interrupted at any instant in time and have its execution resumed later.

\item {\em Identical} $\CPU$s: all the $\CPU$s have the same computing capacities. 
\end{itemize}

Each task $\tau_i$ is a periodic constrained-deadline task characterized by four parameters $(O_i, C_i, D_i, T_i)$ where $O_i$ is the first release time (offset), $C_i$ is the Worst Case Execution Time (WCET), $D_i \le T_i$ is the relative deadline and $T_i$ is the period, i.e., the {\em exact} inter-arrival time between two consecutive releases of task $\tau_i$. These parameters are given with the interpretation that task $\tau_i$ generates an infinite number of successive jobs $\tau_{i,j}$ from time instant $O_i$, with execution requirement of at most $C_i$ each, the $j^{th}$ job which is released at time $O_{i,j}~\equals~O_i~+~(j~-1) \cdot T_i$ must complete within $[O_{i,j}, d_{i,j})$ where $d_{i,j} \equals O_{i,j} + D_i$, the absolute deadline of job $\tau_{i,j}$. 

\drop{Figure~\ref{fig:model} illustrates the task model.

\begin{figure}[!h]
\centering
\includegraphics[width=\linewidth]{modelper.pdf}
\caption{Task model}
\label{fig:model}
\end{figure}
}

We assume without any loss of generality that $O_i \ge 0, \:\: \forall i \in \{1, 2, \ldots, n\}$ and we denote by $O_{\max}$ the maximal value among all task offsets, i.e., $O_{\max} \equals \max\{O_1, O_2, \ldots, O_n\}$. We denote by $P$ the {\em hyperperiod} of the system, i.e., the least common multiple ($\lcm$) of all tasks periods: $P~\equals~\lcm\{T_1, T_2, \ldots, T_n\}$. Also, we denote by $C_{\tau}$ the sum of the $\WCET$s of all tasks in $\tau$: $C_{\tau} \equals \sum_{i=1}^n C_i$.

Job $\tau_{i,j}$ is said to be {\em active} at time $t$ if and only if $O_{i,j} \le t$ and $\tau_{i,j}$ is not completed yet. More precisely, an active job is said to be {\em running} at time $t$ if it has been allocated to a $\CPU$ and is being executed. Otherwise, the active job is said to be {\em ready} and is in the ready queue of the operating system. 

We assume that all the tasks are independent, i.e., there is no communication, no precedence constraint and no shared resource (except for the $\CPU$s) between tasks. Also, we assume that any job $\tau_{i, j}$ cannot be executed in parallel, i.e., no job can execute upon more than one $\CPU$ at any instant in time. 

\subsection{Scheduler specifications}
We consider that tasks are scheduled by using the Fixed-Job-Priority ($\FJP$) scheduler $\gEDF$. That is, the following two properties are always satisfied: {}\emph{(i)} the shorter the absolute deadline of a job the higher its priority and \emph{(ii)} a job may begin execution on any $\CPU$ and a preempted job may resume execution on the same $\CPU$ as, or a different $\CPU$ from, the one it had been executing on prior to preemption. We assume in this research  that the preemptions and migrations of all tasks and jobs in the system are allowed at no cost or penalty.

\section{Definitions and properties}\label{Definitions and properties}

In this section we provide definitions and properties that will help us establishing our exact schedulability test.
First, we formalize the notions of {\em synchronous} and {\em asynchronous} systems, {\em schedule} and {\em valid schedule}, and  {\em configuration}.

\begin{definition}[(A)Synchronous systems]
A task system $\tau = \{\tau_1, \tau_2, \ldots, \tau_{n}\}$ is said to be {\em synchronous} if each task in $\tau$ has its first job released at the same time-instant $c$, i.e., $O_i = c$ for all $1 \le i \le n$. Otherwise, $\tau$ is said to be {\em asynchronous}.
\end{definition}

\begin{definition}[Schedule $\sigma(t)$]
For any task system $\tau = \{\tau_1, \tau_2, \ldots, \tau_{n}\}$ and any set of $m$ identical $\CPU$s $\{\pi_1, \pi_2, \ldots, \pi_m\}$, the schedule $\sigma(t)$ of system $\tau$ at time-instant $t$ is defined as $\sigma: \mathbb{N} \rightarrow \{1, 2, \ldots, n\}^m$ where $\sigma(t) \equals (\sigma_1(t), \sigma_2(t), \ldots, \sigma_m(t))$ with
\[
\sigma_j(t) \equals
  \left\{
          \begin{array}{ll}
	 0,  & \text{if there is no task scheduled on } \pi_j \text{ at time-instant } t\\
          i,   & \text{if task } \tau_i \text{ is scheduled on } \pi_j 
              \text{ at time-instant } t.       	
          \end{array}
 \right.
 \]
\end{definition}

\begin{definition}[Valid schedule] A schedule $\sigma$ of a task system $\tau = \{\tau_1, \tau_2, \ldots, \tau_{n}\}$ is said to be {\em valid} if and only if no task in $\tau$ ever misses a deadline when tasks are released at their specified released times.
\end{definition}

\begin{definition}[Configuration $C_\mathcal{S} (\tau, t)$]\label{Configuration}
Let $\mathcal{S}$ be the schedule of a task system $\tau$. We define the {\em configuration} of the schedule $\mathcal{S}$ at time $t$, denoted by $C_\mathcal{S} (\tau, t)$, to be the $n$-tuple $(e_{1, t}, e_{2, t}, \ldots, e_{n, t})$, where $e_{i, t}$ is the amount of time for which task $\tau_i$ has executed since its last release time up until time $t$, and $e_{i, t}$ is undefined if $t < O_i$. In the latter case, $C_\mathcal{S} (\tau, t)$ is undefined.
\end{definition}

Following Definition~\ref{Configuration}, the configuration $C_\mathcal{S} (\tau, t)$ at time $t$ of a schedule $\mathcal{S}$ is defined if and only if $t \ge O_{\max}$. Moreover, we have $0 \le e_{i, t} \le C_i, \:\: \forall t \ge 0$. Now let $t \ge O_{\max}$ and $t' \ge O_{\max}$ be two time instants such that $t \neq t'$, we denote by $C_\mathcal{S} (\tau, t) \succeq C_\mathcal{S} (\tau, t')$ the fact that $e_{i, t} \ge e_{i, t'}, \forall 1 \le i \le n$.

From now on, we always assume an implementation of $\gEDF$ which is {\em deterministic}, {\em work-conserving} and {\em request-dependent}~\cite{CG06} according to the following definitions.

\begin{definition}[Deterministic schedulers]
A scheduler is said to be {\em deterministic} if and only if it generates a unique schedule for any given set of jobs.
\end{definition}

\begin{definition}[Work-conserving schedulers]
A~scheduler is said to be {\em work-conserving} if and only if it never idles a $\CPU$ while there is at least one active ready task.
\end{definition}

\begin{definition}[Request-dependent schedulers]\label{request-dep}
A scheduler is said to be {\em request-dependent} if and only if for any two tasks $\tau_i$, $\tau_j \in \tau$ and any two jobs $\tau_{i,k}$, $\tau_{j,\ell}$ such that $\tau_{i,k}$ is assigned a higher priority than $\tau_{j,\ell}$, then we also have that $\tau_{i,k + P/T_i}$ is assigned a higher priority than $\tau_{j,\ell + P/T_j}$.
\end{definition}

Informally speaking, Definition~\ref{request-dep} requires that the very same total order is used each hyperperiod between the ``corresponding'' jobs in terms of priorities.

The deterministic and request-dependent requirements are mandatory to ensure a periodic schedule and these requirements impact, eventually, on the $\gEDF$ tie-breaker, i.e., the tie-breaker must be deterministic and request-dependent.

Some further definitions.

\begin{definition}[$\mathcal{A}$-feasibility]
A periodic constrained-deadline task system $\tau$ is said to be {\em $\mathcal{A}$-schedulable} upon a set of $m$ identical $\CPU$s if all the tasks in $\tau$ meet all their deadlines when scheduled using scheduler $\mathcal{A}$, i.e., scheduler $\mathcal{A}$ produces a valid schedule.
\end{definition}

\begin{definition}[Predictability]
A scheduler $\mathcal{A}$ is said to be {\em predictable} if the $\mathcal{A}$-feasibility of a set of tasks implies the $\mathcal{A}$-feasibility of another set of tasks with identical release times and deadlines, but smaller execution requirements.
\end{definition}

Before we present the main result of this paper, we need to introduce the following notations and results taken from~\cite{Ha1994Validating-timi} and~\cite{CG06}.

\begin{lemma}[Ha and Liu \cite{Ha1994Validating-timi}]\label{predictability} 
Any~work-conserving and~$\FJP$~scheduler is {\em predictable} upon identical multiprocessor platforms.
\end{lemma}

Thanks to Lemma~\ref{predictability}, we are guaranteed that the $\gEDF$ scheduler is predictable. Indeed, $\gEDF$ is a work-conserving and~$\FJP$~scheduler. Thereby, given a periodic constrained-deadline task system $\tau$, we can always assume an instance of $\tau$ in which all jobs execute for their whole $\WCET$s. This leads us to consider hereafter a system having known jobs release times, deadlines and execution times. If $\mathcal{S}^{\operatorname{worst}}$ is the valid schedule obtained with these parameters by using the $\gEDF$ scheduler, then we are guaranteed to successfully schedule every other possible instance of $\tau$ in which jobs can execute for less than their $\WCET$s by using the same $\gEDF$ scheduler. 

\begin{lemma}[Cucu and Goossens \cite{CG06}]\label{perioridicity} 
Let $\mathcal{S}$ be the schedule of a periodic constrained-deadline task system $\tau$ constructed by using the $\gEDF$ scheduler. If the deadlines of all task computations are met, then $\mathcal{S}$ is {\em periodic} from some point with a period equal to $P$. 
\end{lemma}

\begin{lemma}[Inspired from Cucu and Goossens~\cite{CG07}]\label{monotony} 
Let $\mathcal{S}$ be the schedule of a periodic constrained-deadline task system $\tau$ constructed by using the $\gEDF$ scheduler. Then, for each task $\tau_i$ and for each time instant $t_1 \ge O_i$, we have $e_{i, t_1} \ge e_{i, t_2}$, where $t_2 \equals t_1 + P$. 
\end{lemma}

\begin{proof}
The proof is made by contradiction. We assume there is some task $\tau_{j_1}$ and some time instant $t_1 \ge O_{j_1}$ such that $e_{j_1, t_1} < e_{j_1, t_2}$, where $t_2 = t_1 + P$. Then there must be some time instant $t'_1 < t_1$ such that $\tau_{j_1}$ is active at both $t'_1$ and $t'_2 = t'_1 + P$, and $\tau_{j_1}$ is scheduled at $t'_2$ while is not at $t'_1$. This can only occur is there is another task $\tau_{j_2}$, which is active (running) at $t'_1$ but not at $t'_2$. But this implies that $e_{j_2, t'_1} < e_{j_2, t'_2}$. Thus, we may repeat the above argument to produce an infinite progression of tasks $\tau_{j_3}$, $\tau_{j_4}$, \ldots, for which no lower bound will exist for the time at which the tasks in the sequence are active. But this is impossible, since every task $\tau_i$ in $\tau$ has an initial release time $O_i$ and the time is discrete in our model of computation. The lemma follows.
\end{proof}

Note that for any task $\tau_i \in \tau$ and for any time instant $t~\ge~O_{\max}$, $C_\mathcal{S} (\tau, t)$ is {\em monotonically decreasing} relative to $t$ with period $P$ thanks to Lemma~\ref{monotony}, i.e., 
$C_\mathcal{S} (\tau, t + k \cdot P) \succeq C_\mathcal{S} (\tau, t + (k+1) \cdot P) \quad \forall k \in \mathbb{N}$

\section{Exact schedulability test}\label{The proposed approach}

In this section we provide an exact schedulability test for the $\gEDF$ scheduling of periodic hard real-time tasks upon identical multiprocessor platforms. It is worth noticing that we assume in this section that each job of the same task (say $\tau_i$) has an execution requirement which is exactly $C_i$ time units thanks to the predictability property of this scheduler. Based on the later result, the intuitive idea behind our approach is to construct a schedule by using an implementation of $\gEDF$ which follows hypothesis described in Section~\ref{Definitions and properties}, then check to see if the deadlines of all task computations are met. However, for this method to work we need to establish an ``a priori'' time interval within which we need to construct the schedule. If the task system $\tau$ is synchronous, then such a  time interval is known: $[0, P)$ where $P = \lcm\{T_1, T_2, \ldots, T_n\}$ see~\cite{CG06} for details. Unfortunately, if the task system $\tau$ is asynchronous, such a time interval is unknown, in the following we will fill the gap.

As the task system $\tau$ is composed of periodic tasks, the idea thereby consists in simulating the system until the schedule becomes periodic, i.e., the {\em steady phase} representing the general timely behavior of the system from a certain time instant is reached. This steady phase is reached when two configurations separated by $P$ time units are identical. \\

Before going any further in this paper, it is worth noticing the following interesting observations.

\paragraph{Observation~1.}\label{Observation1} By extending the results obtained in the uniprocessor framework to the multiprocessor platforms, Leung claimed in~\cite{Leung89} that an exact feasibility condition for $\gEDF$ consists in checking if (i) every deadline is met until time $O_{\max} + 2P$ and (ii) the configurations at instants $O_{\max} + P$ and $O_{\max}+2P$ are identical. Anyway, this is flaw, since there are schedulable task systems that reach their steady phase \emph{later} than $O_{\max} + 2P$, as shown by Counterexample~\ref{counterexample1} taken from~\cite{BC07}.

\begin{counterexample}[Braun and Cucu \cite{BC07}]\label{counterexample1} Consider the following periodic task system: $\tau_{1} = (O_1=0, C_1=2, D_2 = T_2 = 3), \tau_{2} = (O_2=4, C_2=3, D_2 = T_2 = 4), \tau_{3} = (O_3=1, C_3=3, D_3 = T_3 = 6)$ to be scheduled with $\gEDF$ upon $m = 2$ $\CPU$s.

\drop{
\begin{table}[h]
\begin{center}
\begin{tabular}{|c|c|c|c|}
\hline
Tasks        & $O_i$ & $C_i$ & $D_i = T_i$ \\
\hline
$\tau_1$  &   $0$   &    $2$   &   $3$      \\
\hline
$\tau_2$  &   $4$   &    $3$   &   $4$     \\
\hline
$\tau_3$  &   $1$   &    $3$   &   $6$     \\
\hline
\end{tabular}
\label{taskparam1}
\caption{Task parameters}
\end{center}
\end{table}
}

By building the schedule (see Figure~\ref{fig:ce1}), it is possible to see that at time $O_{\max} + P = 4+12 = 16$ and $O_{\max} + 2P = 4+ 2 \cdot 12 = 28$ the steady phase has not yet been reached (at times $17$ and $29$ there are two different configurations). However, the steady phase is reached after a further hyperperiod. Since no deadline is missed, the task system is schedulable with $\gEDF$.
\begin{figure*}
\begin{center}
	\includegraphics[width=0.66\linewidth]{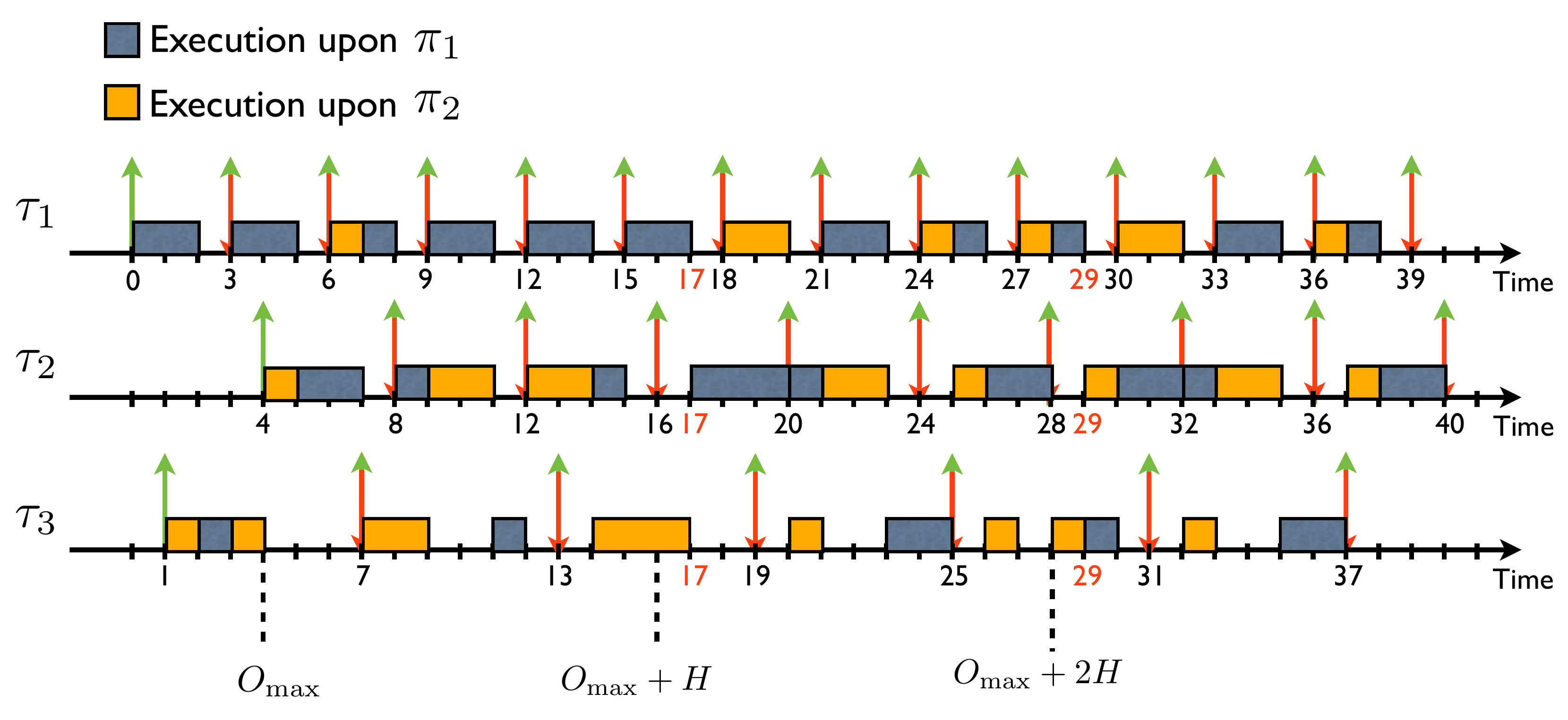}
	\caption{$[0, O_{\max} + 2P]$ is not a feasibility interval for the $\gEDF$ scheduler.}
	\label{fig:ce1}
\end{center}
\end{figure*}
\end{counterexample}

The flaws in the results proposed by Leung in~\cite{Leung89} come from many sources. The paper was actually centered on the Least Laxity First ($\LLF$) scheduler defined as follows.

\begin{definition}[$\LLF$ scheduler]
The $\LLF$ scheduler always executes the jobs with least laxity; being the laxity of a job its absolute deadline minus the sum of its remaining processing time and the current time.
\end{definition}

$\triangleright$ {\bf Flaw~1.} The first flaw is about the comparison between the $\LLF$ scheduler and the $\gEDF$ scheduler. Indeed it was claimed in~\cite{Leung89} that every instance of jobs schedulable by $\gEDF$ upon $m$ $\CPU$s is also schedulable by $\LLF$ upon $m$ $\CPU$s. This flaw, presented as one of the main results of the paper, has been pointed out by Kalyanasundaram et al. in~\cite{KKT00}. \\

$\triangleright$ {\bf Flaw~2.} The second flaw lies in the proof of Lemma~2 of Leung's paper replicated here (except we consider $\gEDF$).

\begin{lemma}[Lemma~2, pages 216--217 of~\cite{Leung89}]\label{lem:leung}
Let {\cal S} be the schedule of a task system $\tau$ constructed by using the $\gEDF$ scheduler upon $m \ge 1$ $\CPU$s. If $\tau$ is schedulable by using the $\gEDF$ scheduler on $m$ $\CPU$s, then $C_\mathcal{S} (\tau, t_1) = C_\mathcal{S} (\tau, t_2)$ where $t_1 \equals O_{\max} + P$ and $t_2 \equals t_1 + P$.
\end{lemma}

The argument stating that: 
\begin{quote} 
the $m$ $\CPU$s are always busy in the interval $[t_1, t_2]$
\end{quote}
 is incorrect; this is a \emph{uni}processor argument not valid in a multiprocessor context. Indeed, considering Conterexample~\ref{counterexample1}, it is not difficult to see in Figure~\ref{fig:ce1} that $t_1 = 16$, $t_2 = 28$ and $C_\mathcal{S} (\tau, 16) \neq C_\mathcal{S} (\tau, 28)$. However, in the time-slots $[17, 18)$ and $[23, 24)$, only one CPU (here, $\CPU$ $\pi_1$) out of two is actually busy by the execution of the jobs.

\paragraph{Observation~2.}\label{Observation2}  Although we address the $\gEDF$ scheduling problem of periodic constrained-deadline task systems,~Counterexample~\ref{counterexample2} give evidence of possibly late occurrence of the {\em steady phase} in the valid schedule $\mathcal{S}$ of a task system $\tau$. Indeed, it shows that the steady phase can be reached after a time-instant as large as $O_{\max} + {\bf 42} \cdot P$.
  
\begin{counterexample}\label{counterexample2} 
Consider the following periodic task system: $\tau_{1} = (O_1=225, C_1=90, D_2 = T_2 = 161), \tau_{2} = (O_2=115, C_2=40, D_2 = T_2 = 161), \tau_{3} = (O_3=0, C_3=72, D_3 = T_3 = 161), \tau_{4} = (O_4=129, C_4=120, D_4 = T_4 = 161)$ to be scheduled with $\gEDF$ upon $m = 2$ $\CPU$s.

\drop{
\begin{table}[h]
\begin{center}
\begin{tabular}{|c|c|c|c|}
\hline
Tasks        & $O_i$ & $C_i$ & $D_i = T_i$ \\
\hline
$\tau_1$  &   $225$   &   $90$   &   $161$    \\
\hline
$\tau_2$  &   $115$   &   $40$   &   $161$    \\
\hline
$\tau_3$  &   $0$   &    $72$   &   $161$       \\
\hline
$\tau_4$  &   $129$   &    $120$   &   $161$   \\
\hline
\end{tabular}
\end{center}
\label{taskparam2}
\caption{Task parameters}
\end{table}
}
By building the schedule using an open source simulation tool such as $\operatorname{STORM}$\footnote{$\operatorname{STORM}$ stands for ``Simulation Tool for Real-time Multiprocessor Scheduling Evaluation'' and is a simulation tool developed at Irccyn, \'Ecole Centrale de Nantes, France.} (we implemented a deterministic and request-dependent EDF tie-breaker), it is possible to see that at time-instants $O_{\max} + 42P = 6987$ and $O_{\max} + 43P = 7148$ the steady phase has not been reached yet (there are two different configurations at time-instants $6988$ and $7149$). However, the steady phase is reached after a further hyperperiod. Again, since no deadline is missed, the task system is schedulable with $\gEDF$.
\end{counterexample}

It thus follows from Lemma~\ref{perioridicity}, Observation~$1$ and Observation~$2$ the conjecture that integer $k \in \mathbb{N}^{+}$ in Expression~$(O_{\max} + k \cdot P)$ for the time-instant to reach the steady phase must be a function of tasks parameters.

\begin{lemma}\label{lem:bound}
Let $\mathcal{S}$ be the valid schedule of an asynchronous periodic constrained-deadline task system $\tau$. We assume that $\mathcal{S}$ has been constructed by using the $\gEDF$ scheduler. Then an upper bound on the time-instant up to which there exists $t > 0$ such that $C_\mathcal{S} (\tau, t-P) = C_\mathcal{S} (\tau, t)$ is given by $t_{\operatorname{up}}\equals O_{\max} + \left(C_{\tau} + 1\right) \cdot P$
\end{lemma}

\begin{proof}
Let $\mathcal{S}$ be the valid schedule of an asynchronous periodic constrained-deadline task system $\tau$, constructed by using the $\gEDF$ scheduler. By definition, $O_{\max}$ is the first time-instant at which all tasks in $\tau$ are released, the steady state cannot start before that time-instant. Now we will upper-bound the first time-instant where the schedule starts to repeat. The worst-case scenario is the one where there is a \emph{different} system configuration, between two successive hyperperiods, a \emph{maximal} number of times starting from $O_{\max}$. I.e., $e_{i, O_{\max}}=C_{i} \quad\forall i$ and $C_\mathcal{S} (\tau, O_{\max} + k \cdot P) \neq C_\mathcal{S} (\tau, O_{\max} + (k + 1) \cdot P) \quad  \forall k \in [0, \hat{k}]$ and 
$C_\mathcal{S} (\tau, O_{\max} + \hat{k} \cdot P) = C_\mathcal{S} (\tau, O_{\max} + (\hat{k} + 1) \cdot P)$.
By Lemma~\ref{monotony} and the fact that time is discrete in our model of computation, the worst-case scenario ---i.e., the scenario which maximizes $\hat{k}$--- corresponds to the case where each time the system configuration differs from the previous one in the following way: $\exists \ell \in [1,n]$ such that: $e_{i, O_{\max} + (k+1) \cdot P} = e_{i, O_{\max} + k \cdot P} \quad \forall i \neq \ell$ and $e_{\ell, O_{\max} + (k+1) \cdot P} = e_{\ell, O_{\max} + k \cdot P} - 1$. Since $0 \leq e_{i,t} \leq C_{i}$ an upper-bound for $\hat{k}$ is $\sum_{i=1}^{n} C_{i}$. And by Lemma~\ref{perioridicity} at time $t_{\operatorname{up}} = O_{\max} + \left(\sum_{i=1}^n C_i + 1\right) \cdot P = O_{\max} + \left(C_{\tau} + 1\right) \cdot P$ we have $C_\mathcal{S} (\tau, t_{\operatorname{up}}-P) = C_\mathcal{S} (\tau, t_{\operatorname{up}})$ and the schedule is periodic with a period of $P$.
\end{proof}

Now we have the material to define an ``exact'' schedulability test for the $\gEDF$ scheduling of periodic hard real-time tasks upon identical multiprocessor platforms.

\paragraph{Exact Schedulability Test.} Let $\tau$ be an asynchronous periodic constrained-deadline task system. Let $\pi$ be a platform consisting of $m$ {\em identical} $\CPU$s. We assume that $\tau$ is scheduled upon $\pi$ by using the $\gEDF$ scheduler. Then $\tau$ is schedulable if and only if 
\emph{(i)} all deadlines are met in $[0, t_{\operatorname{up}})$ and \emph{(ii)} $C_\mathcal{S} (\tau, t_{\operatorname{up}} - P) = C_\mathcal{S} (\tau, t_{\operatorname{up}})$
where $t_{\operatorname{up}}$ is defined in Lemma~\ref{lem:bound}.

\begin{proof}
If there is a deadline miss in $[0, t_{\operatorname{up}})$, then the system is clearly not schedulable. Otherwise, this schedulability test is a direct consequence of Lemma~\ref{lem:bound}, Lemma~\ref{perioridicity} (the periodicity property of the schedule) and Lemma~\ref{predictability} (the predictability property of the $\gEDF$ scheduler).
\end{proof}

\section{Conclusion}\label{Conclusion}
In this paper, we considered the scheduling problem of hard real-time systems composed of periodic constrained-deadline tasks upon identical multiprocessor platforms. We assumed that tasks were scheduled by using the $\gEDF$ scheduler and we provided an {\em exact} schedulability test for this scheduler. Also, we showed by means of a counterexample that the feasibility interval, and thus the schedulability test, proposed by Leung~\cite{Leung89} is incorrect and we showed which arguments are actually incorrect.


\bibliographystyle{latex8}
\bibliography{biblio-feas.bib}
\end{document}